# Entropy-map SSIM analysis of Salt and Pepper Noise Removal via Recursive Median Filterring

Petr Boriskov[a, *] and Andrei Velichko[a]

[a]*Petrozavodsk State University, Petrozavodsk, Russia*
**e-mail: boriskov@petrsu.ru*

[*] Corresponding author

**Abstract** This paper studies the removal of salt-and-pepper (SP) noise from grayscale images using a median filter (MF) within a recursive thresholding algorithm. Denoising performance is assessed using two complementary metrics: SSIM-Img and SSIM-Map. SSIM-Img is standard Image Quality Assessment (IQA), the conventional Structural Similarity Index (SSIM) computed between the restored and clean images. SSIM-Map is novel IQA, an SSIM-based evaluation computed between entropy maps of these images, where the maps are obtained using singular value-decomposition entropy in sliding windows. We show that SSIM-Map is more sensitive to residual impulse artifacts, blur, and local intensity transitions, and therefore complements the conventional SSIM-Img metric. Experiments on grayscale images demonstrate that recursive median filtering with thresholding can restore images corrupted by high-density SP noise, including noise levels of 50–60%. We also propose a two-scale scheme, denoted 2-SRMAT, which combines two median filters with different window sizes and a final thresholding step. The proposed scheme is simple, requires no training, and is potentially suitable for resource-constrained edge/IoT implementations. The 2-SRMAT is not a universal replacement for recent learning-based or optimization-based denoisers; rather, it is used as an restoration framework for demonstrating the practical value of entropy-map-based structural assessment. The results confirm that new entropy-map-based IQA can provide an additional objective assessment of residual defects and blur after impulse-noise removal.

## 1 Introduction

Over the past decade, a wide range of image denoising algorithms has been proposed, offering different quality levels and computational costs. Transform-domain methods — wavelet-based [1, 2] and spectral/frequency approaches [3], singular-value–based techniques [1, 4], and three-dimensional filtering [5] — are effective at suppressing noise while preserving useful content. However, these approaches are often complex and require substantial domain expertise due to numerous variants and parameters. For example, the choice of the mother wavelet and the decomposition level strongly affects performance. In addition, deploying computationally demanding algorithms on low-power or embedded platforms can be challenging.

Local filters such as the moving-average and the MF are long-established, offering very high throughput at low computational cost. In their simplest forms, though, they provide limited tuning — essentially only the sliding-window (kernel) size. As the kernel grows, stronger denoising is accompanied by increased blurring and reduced contrast.

Impulse noise — often called salt-and-pepper (SP) noise in images — remains a major challenge in digital image processing [6, 7]. Among common filters, the MF is one of the most effective for SP noise removal [6-8]. The MF is a nonlinear operator: pixels within each sliding window are ranked, and the central order statistic is returned. Unlike the mean filter, it suppresses extreme outliers without greatly affecting neighboring values. Because edges are largely preserved, the median filter can be applied recursively. The advent of recursive MFs [8,9] further improved the denoising efficacy of median-based schemes. To mitigate blurring, the MF is often combined with thresholding [10]: the filtered image is compared with the original, and original pixel values are retained where a specified threshold condition is satisfied.

Entropy, as a measure of order (disorder) in systems, is widely used in information technology, including image processing and analysis, such as image fusion [11, 12], segmentation [13, 14], recognition [15], etc. Adaptation algorithms on the basis of entropy [16, 17], including MF [18], are also developed. In this work, as the quality metric of image restoring, we use the standard SSIM metric [19]. However, the key novelty of our approach is the systematic use of SSIM not only when comparing images, but also their entropy maps, which we obtain by calculating the singular value decomposition entropy (SVDEn) [20, 21] over sliding windows. The main methodological emphasis is placed on the entropy-map SSIM analysis of residual artifacts, blur, and local structural loss for simple (training-free) MF-based restoration.

We perform a detailed SSIM-based analysis of recursive thresholded median filtering for the removal of SP corruption from grayscale images. In the proposed recursive scheme, a threshold rule is inserted into the output-to-input feedback loop of the MF: output pixels are either retained or replaced with the corresponding pixels from the original noisy image. Building on this recursive procedure, we also propose a two-scale recursive median algorithm with thresholding, denoted 2-SRMAT, which combines median filters with different sliding-window sizes. In all experiments considered in this work, the corrupted image is used as the input, whereas the clean reference image is used only for quantitative evaluation of the denoising quality.

Main contributions:

– We present SSIM-Map as new IQA, an SSIM metric computed between image entropy maps (SVDEn in sliding windows) that complements the standard SSIM for images and serves as a sensitive indicator of residual impulse artifacts, over-smoothing, and local detail degradation;

– We analyze recursive thresholded median filtering for SP-corrupted images and show how the choice of threshold, recursion number, and median-filter window size affects the restoration quality;

– We propose a simple two-scale denoising scheme (2-SRMAT) that combines small- and large window median filtering with a final thresholding step, and we position it as a transparent, training-free simplicity–quality trade-off.

The rest of the paper is organized as follows. Section 2 describes the recursive denoising scheme with a threshold rule, and the SSIM computation for images and their entropy maps. Section 3 (Results) presents the results of restoration of the grayscale image of Lena (512×512) px and a number of other images (see Fig. 14 and Table 5), including noise reduction using the 2-SRMAT scheme in Sections 3.2. Sections 4 (Discussion) and 5 (Conclusion) summarize the proposed algorithms. All computations were performed in Python.

## 2 Methods

### 2.1 SP denoising by MF with recursive thresholding

The general MF scheme with recursive thresholding is shown in Fig. 1, where the recursing means that the once-filtered image is fed back to the input. The initial noisy image is fed to the MF input with SP corruption and normalized grayscale intensity in the range [0, 1] with 0 (black) and 1 (white). SP noise is modeled as uncorrelated defects of individual pixels, randomly scattered across a reference image grid with values equal to 1 (“salt”) or 0 (“pepper”). The SP noise level $\mu_{sp}$ in this model, called fixed [6], is defined as the proportion of damaged pixels:

$$\mu_{sp}(\%) = 100 \cdot \frac{n_c}{n}, \quad (1)$$

where $n_c$ is the combined number of corrupted "salt" and "pepper" pixels (assumed equal density and the selected corrupted pixels were sampled without replacement), and $n$ is the total number of pixels in the image.

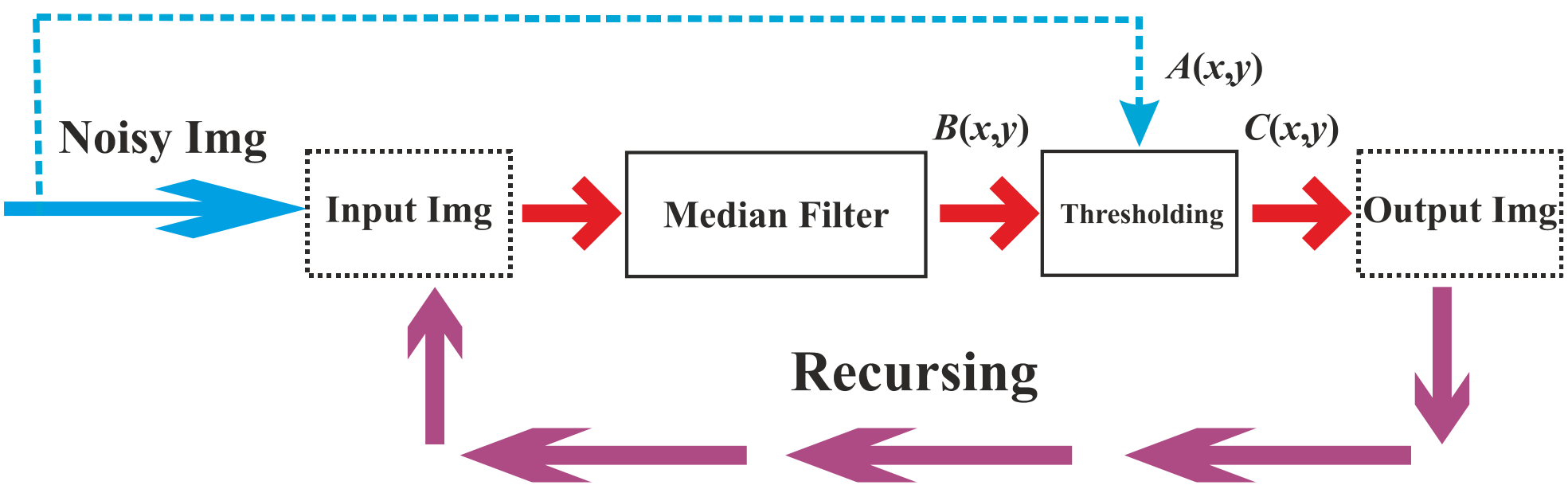

**Fig. 1.** General scheme of median filter with recursive thresholding. The noisy image (Noisy Img) is used only at the first pass (blue solid arrow) and in thresholding (2) (blue dashed arrow). $A(x, y)$, $B(x, y)$, and $C(x, y)$ denote pixels of the noisy, MF output, and post-threshold images, respectively. The recursing between the input (Input Img) and output (Output Img) images is shown

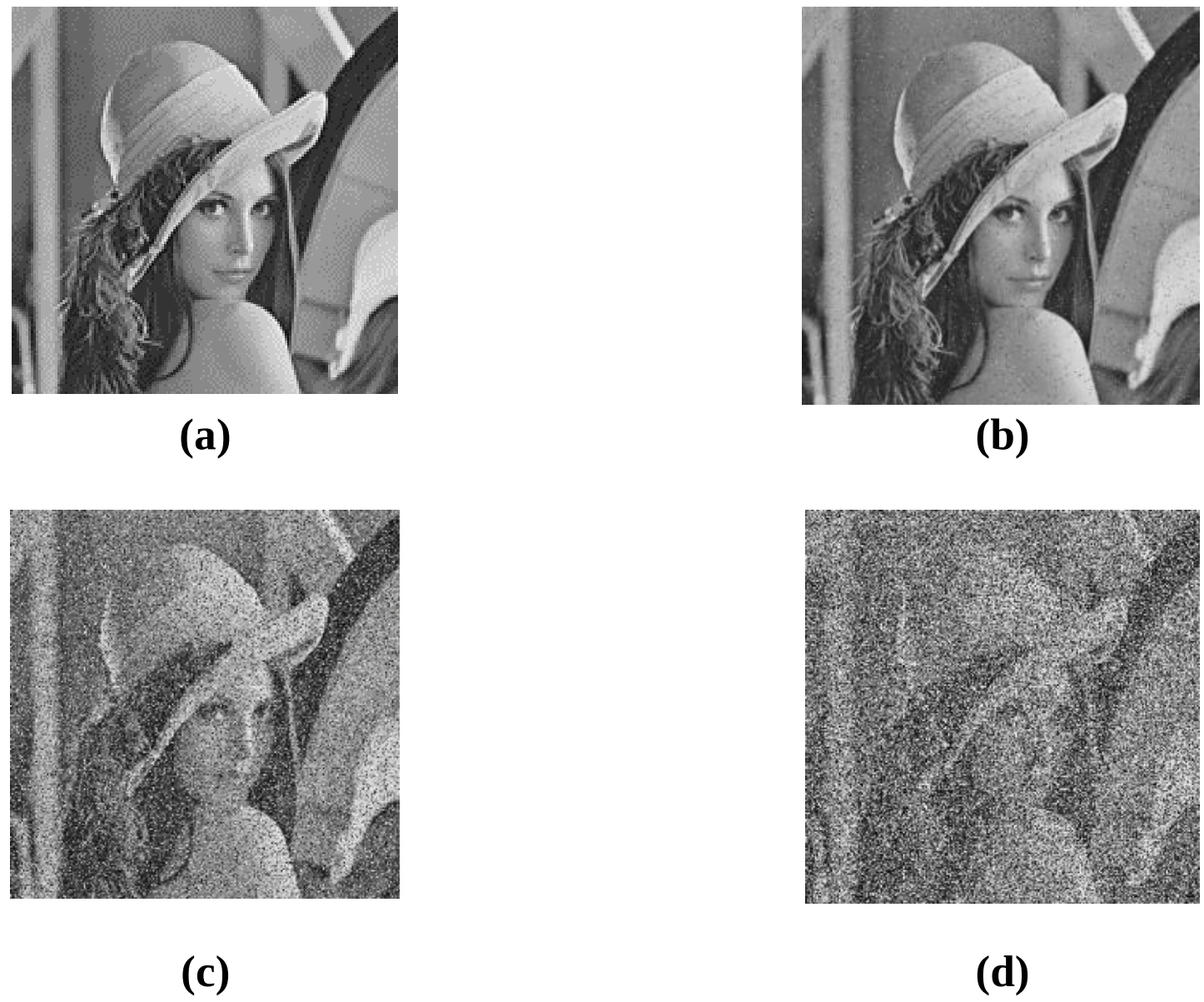

**Fig.2.** Examples of grayscale images of Lena (512×512) px: clean (a) and noisy versions with $\mu_{sp}$ = 2 % (b), $\mu_{sp}$ = 30 % (c), and $\mu_{sp}$ = 60 % (d)

**Threshold rule.** In both DnF scenarios, a threshold rule is inserted into the output-to-input feedback loop: depending on this rule, output pixels are either retained or replaced by the corresponding pixels from the noisy image. For every pixel the thresholding is defined as

$$C(x,y)=\begin{cases} A(x,y), \ if \ |A(x,y)-B(x,y)|<threshold \\ B(x,y), \ if \ |A(x,y)-B(x,y)|>threshold \end{cases}, \quad (2)$$

where $A(x, y)$, $B(x, y)$ and $C(x, y)$ are grayscale intensities [0,1] at coordinates $(x, y)$ for the noisy, the filter output, and the post-threshold images, respectively (see Fig. 1). Thus, at each DnF pass, output pixels are replaced by the corresponding pixels of initial noisy image when the absolute difference $A(x, y)$ and $B(x, y)$ does not exceed the threshold. Evidently, with a zero threshold the rule degenerates to no thresholding, i.e., $C(x, y) = B(x, y)$ when $threshold = 0$. Equation (2) will be modified for scaled denoising in Sections 3.2.

**MF.** Noise removal is performed using a sliding window method, calculating the median in each window $W$ over the pixels A(x, y) as [7]

$$median = \underset{A(x_o, y_o) \in W}{\arg\min} \sum_{(x,y)} |A(x,y) - A(x_o, y_o)|. \quad (3)$$

Rectangular windows symmetric about the center may be (3×3), (5×5), (7×7), (9×9) px, etc.

## 2.2 SSIM of images and their entropy maps

A standard measure of image similarity is the Structural Similarity Index (SSIM), which compares two images via local-window statistics: means, variances, and covariance of pixel intensities. The combined local metric $\mathrm{SSIM}_{\mathrm{local}}$ is [19]:

$$\mathrm{SSIM}_{\mathrm{local}}(x,y)=\frac{(2\mu_x\mu_y+C_1)(2\sigma_{xy}+C_2)}{(\mu_x^2+\mu_y^2+C_1)(\sigma_x^2+\sigma_y^2+C_2)}, \quad (4)$$

where $\mu_x$ ($\mu_y$) and $\sigma_x^2$ ($\sigma_y^2$) are the means and variances in the corresponding local windows of the two images, $\sigma_{xy}$ is their covariance (the window location is $(x, y)$, $C_1 = (0.01\cdot255)^2$ and $C_2 = (0.03\cdot255)^2$ are empirical constants tied to the intensity range that ensure numerical stability. The global SSIM for an image pair is the arithmetic mean of $\mathrm{SSIM}_{\mathrm{local}}$ over all window positions.

For any raster image, one can define different families of overlapping local windows, sliding with a step of 1, that completely cover the image. For rectangular windows, the number of windows in each family is determined by the window size ($L$ x $L$) and the image size ($M$ x $N$), and is equal to

$$M_L \times N_L = (M - (L-1)) \times (N - (L-1)). \quad (5)$$

Then, for the image, we can construct an entropy map of size (5) by calculating the entropy for each sliding rectangular window and assigning it to some pixel of the window, for example, the upper-left corner pixel. Based on the physical meaning of entropy, its value for a sliding window should be small if the pixels have similar intensities, and large if the pixels differ greatly. Thus, differences in the intensity of image pixels will manifest themselves in changes in the entropy of the nearest sliding windows. One image can have multiple entropy maps depending on the entropy type and parameters, but each map reflects a set of characteristics for only one specific image.

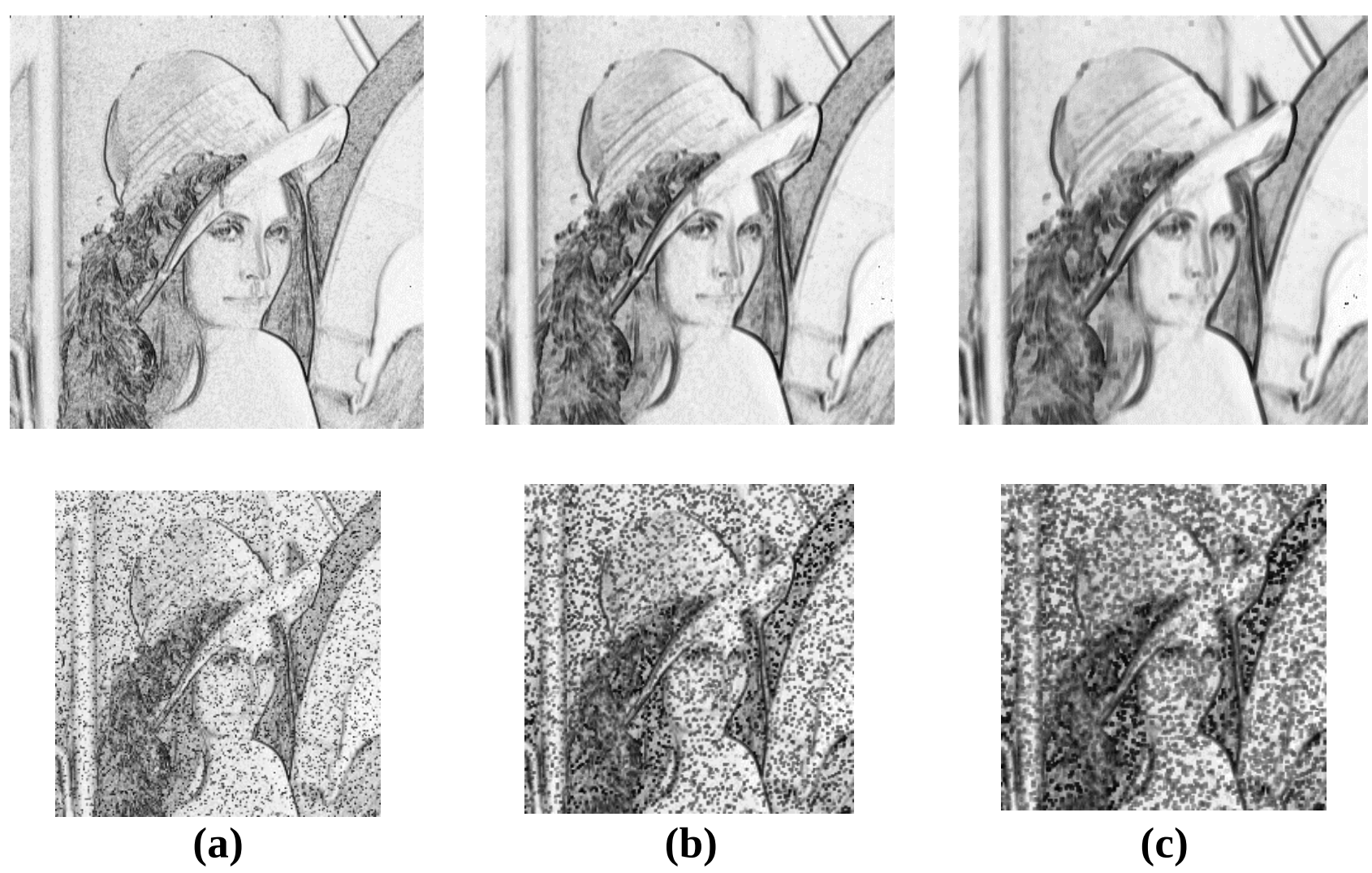

(a) (b) (c)

**Fig. 3.** 8-bit entropy maps of the clean Lena image (top) and the noisy Lena image with $\mu_{sp}$ = 2 % (bottom) computed by SVD entropy with sliding windows in size: (a) – (3×3) px ; (b) – (5×5) px; (c) – (7×7) px

The resulting entropy map is also converted to a grayscale image. Figures 3 illustrate multiple 8-bit ([0,255]) entropy maps of clean (Fig. 2a) and low-noisy (Fig. 2b) image computed using the well-known entropy measure, SVD entropy [20, 21]. For each local window, the SVD entropy is computed from the singular values of the corresponding intensity matrix. Let $\sigma_i$, i = 1,... , $r$ be the non-zero singular values of a local window and they are normalized so that their sum is equal to 1

$$p_i = \frac{\sigma_i}{\sum_{j=1}^{r} \sigma_j} \tag{6}$$

The normalized SVD entropy is then defined as

$$SVDEn = \frac{1}{\log r} \sum_{j=1}^{r} p_j \log p_j \,. \tag{7}$$

This normalization maps the entropy to the interval [0,1], with lower values corresponding to locally homogeneous regions and higher values corresponding to stronger local intensity variation.

From the normalized map (7), it is easy to obtain an 8-bit grayscale map (Fig. 3) such that pixels with low entropy (~ 0) are white (255), and, conversely, pixels with high entropy (~ 1) are black (0). With this grayscale, the entropy maps better match the real image: the girl's face is white, her hair is black. As can be seen in the upper images of Figure 3, the maps vary significantly in the degree of blurring, which increases with the window size. Indeed, the entropy of smaller windows is more sensitive to fine details, which appear during sliding as pixels with sharp changes in intensity. It is also noticeable that clean (Fig. 3a) and low-noise (Fig. 3b) images are virtually indistinguishable, while their entropy maps (Fig. 3) have sharp differences. The noisy image maps (Fig. 3, bottom) show numerous defects compared to the clean image maps (Fig. 3, top), and these defects become more noticeable as the sliding entropy window increases.

When two entropy maps have the same size, their similarity can be assessed exactly as for real images: via local $SSIM_{local}$ (4) and the global SSIM (the average of $SSIM_{local}$ over ll window positions). In the denoising task, we compute entropy maps of the clean and restored images and take the global SSIM as their average over sliding windows. This new parameter is an additional characteristic of the similarity between clean and restored images, which can be used to evaluate the effectiveness of noise reduction. Next, to distinguish metrics, let SSIM-Img denote SSIM for images and SSIM-Map for entropy maps.

# 3 Results

## 3.1 Analysis of SP denoising by MF

Table 1 lists the MF parameters and entropy map parameters of Lena grayscale image. Figure 4 illustrates the effect of the threshold rule (2) on SP denoising in terms of SSIM metrics (SSIM-Img and SSIM-Map). For MF these metrcs are higher with thresholding (Fig. 4b) than without (Fig. 4a). Moreover, as seen from the black curve with squares in Fig. 4a (MF, SSIM-Img), even at very low noise $\mu_{sp} \leq 10$ % the values do not exceed ~ 0.81, whereas the SSIM metrics with thresholding are close to 1 (Fig. 4b). Note that images with SSIM-Img > 0.9 are visually almost indistinguishable from the reference (Fig. 2a), while for SSIM-Img < 0.2 effective restoration is essentially absent.

**Table 1.** MF and entropy map parameters of Lena grayscale image

| Parameters | block shape, $N_{bl}$ | window shape, $N_w$ | Entropy windows size | threshold |
|---|---|---|---|---|
| MF | (512×512) px | (5×5) px | (5×5) px | 0.15 |

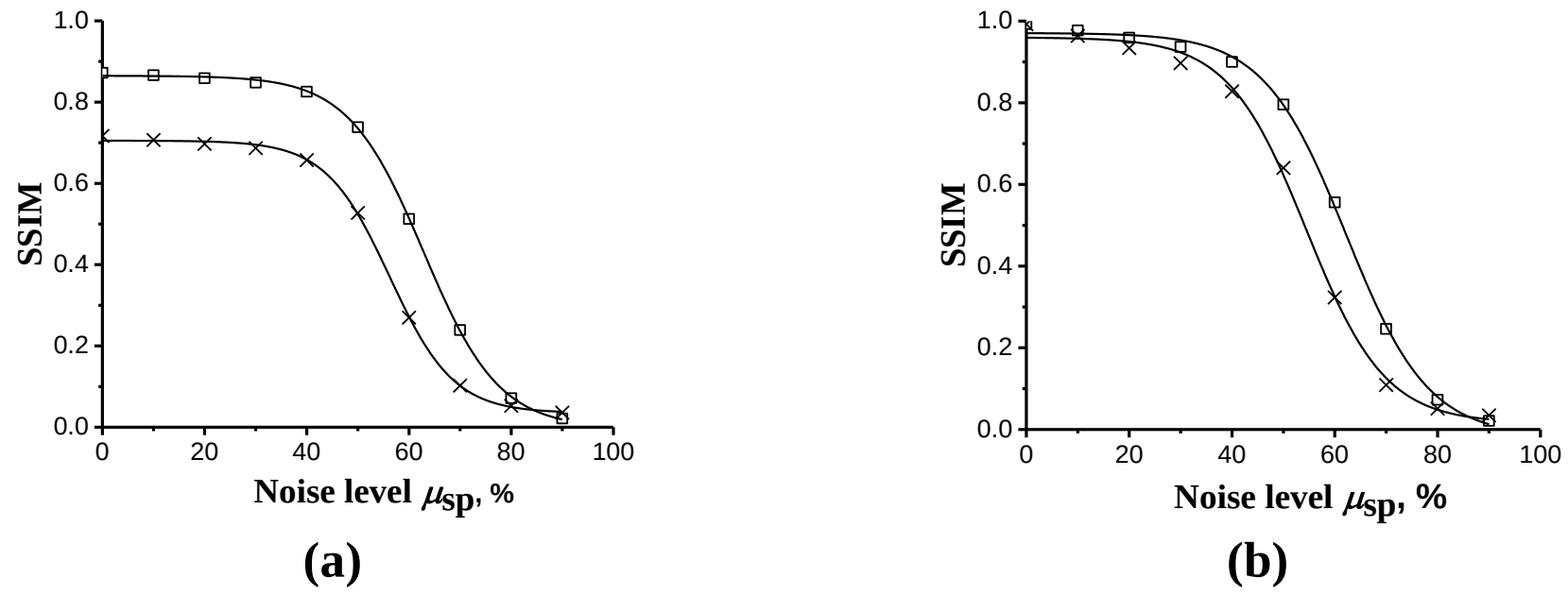

**Fig. 4.** Dependence of SSIM metrics (SSIM-Img – squares and SSIM-Map – crosses) on the SP noise level $\mu_{sp}$ for non-recursive (single-pass) denoising by MF without threshold (a) and with threshold (b). MF and entropy map parameters are in Table 1

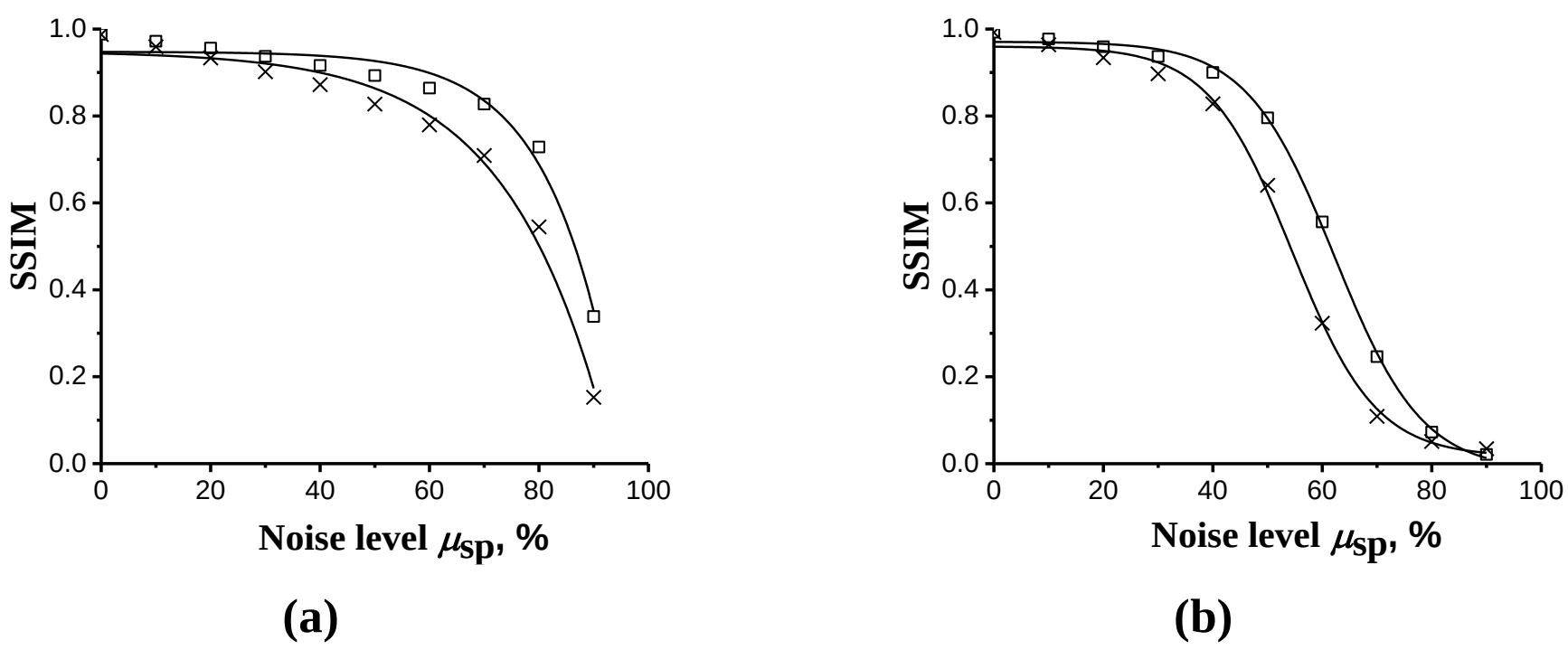

**Fig. 5.** Dependence of SSIM metrics (SSIM-Img – squares and SSIM-Map – crosses) on the SP noise level $\mu_{sp}$ for denoising by MF: after three (a) and ten (b) recursions. MF and entropy map parameters are in Table 1

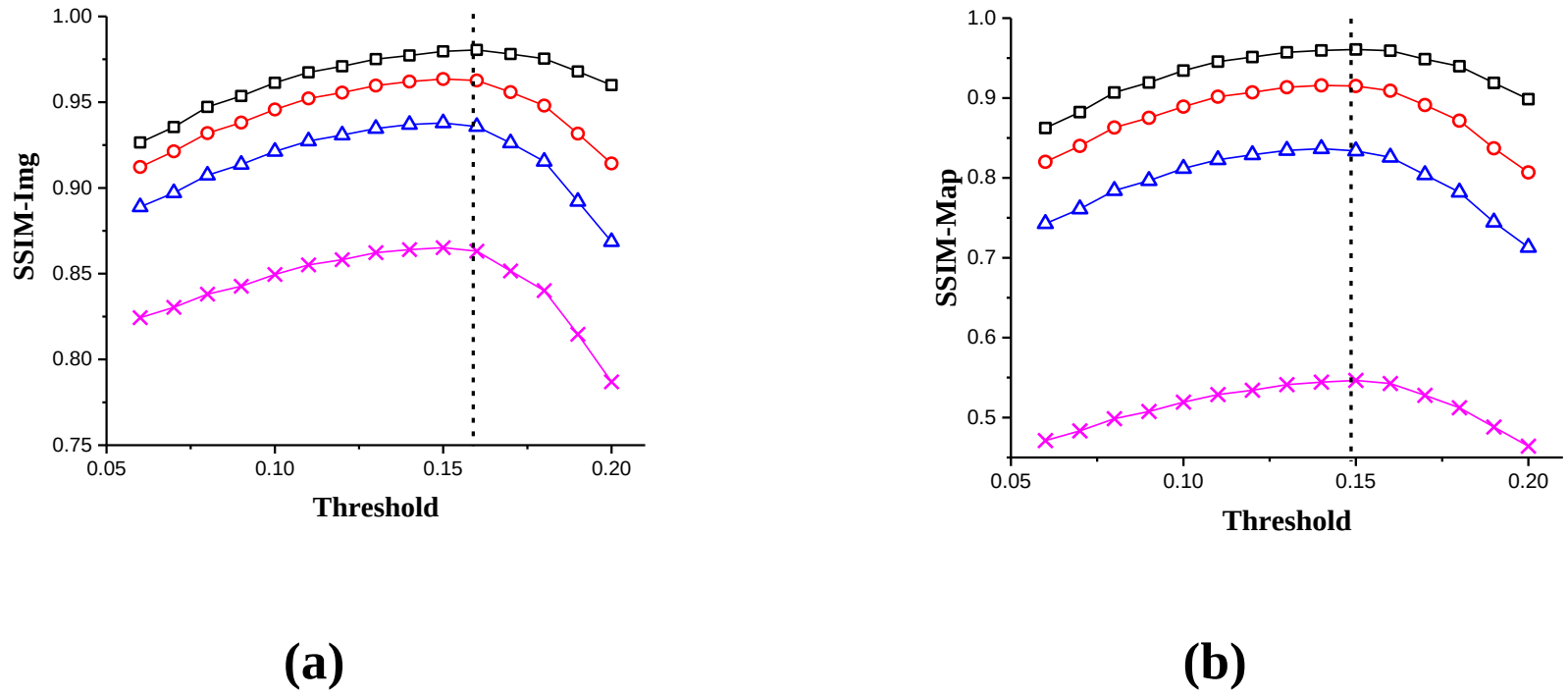

**(a)** **(b)**

**Fig. 6.** Dependence of SSIM-Img (a) and SSIM-Map (b) metrics on the threshold for recursive denoising by MF: $\mu_{sp}$= 5 % (black square), $\mu_{sp}$= 15 % (red circles), $\mu_{sp}$= 30 % (blue rectangles), $\mu_{sp}$= 60 % (magenta crosses). The vertical dotted lines show the approximate maximum of SSIM-Img and SSIM-Map curves. MF and entropy map parameters are in Table 1., the number of recursions is 10

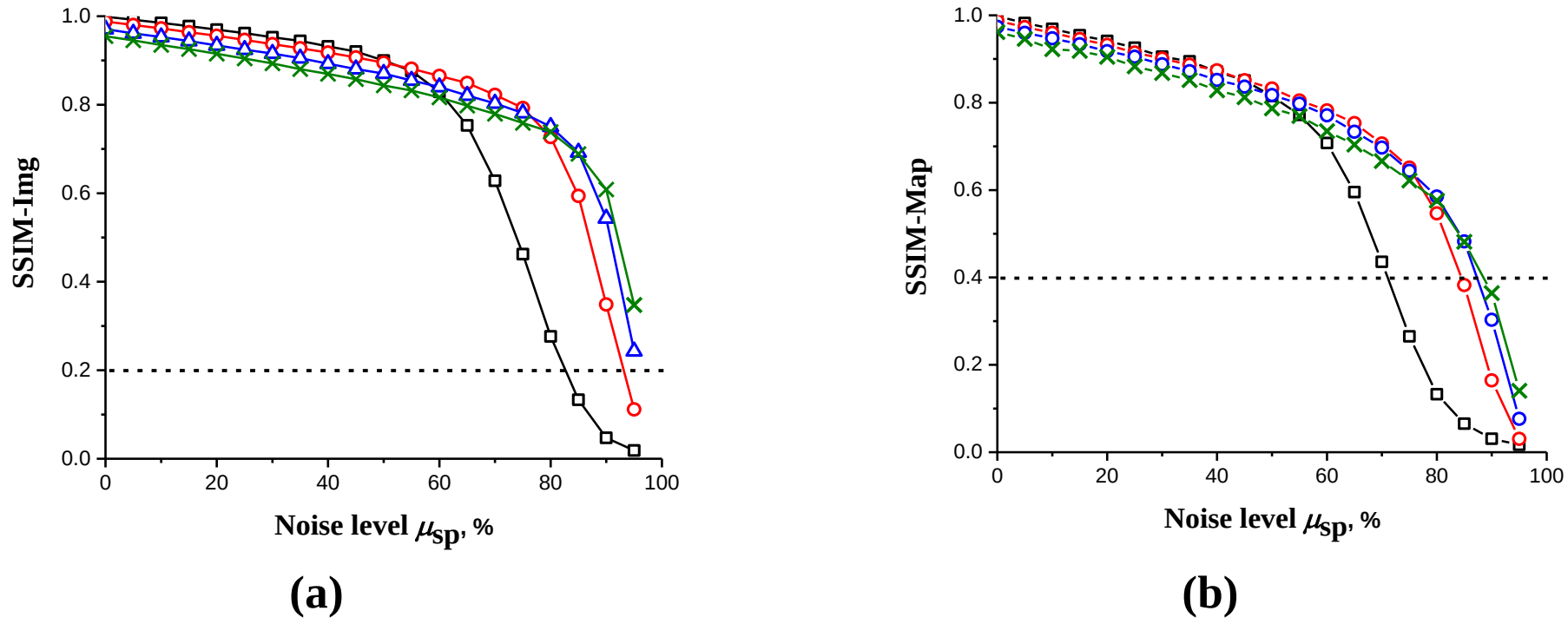

**(a)** **(b)**

**Fig. 7.** Dependence of SSIM-Img (a) and SSIM-Map (b) on the SP noise level $\mu_{sp}$ for recursive denoising by MF with different filter windows size (a): (3×3) px – black squares, (5×5) px – red circles, (7×7) px – blue rectangles, 9×9 px – green crosses. The horizontal dotted lines show the nominal levels below which Lena's true image is almost undiscernible at the restoring image and her entropy map. Entropy windows size (in Fig. 7(b)) is (5×5) px, threshold = 0.15, the number of recursions is 10

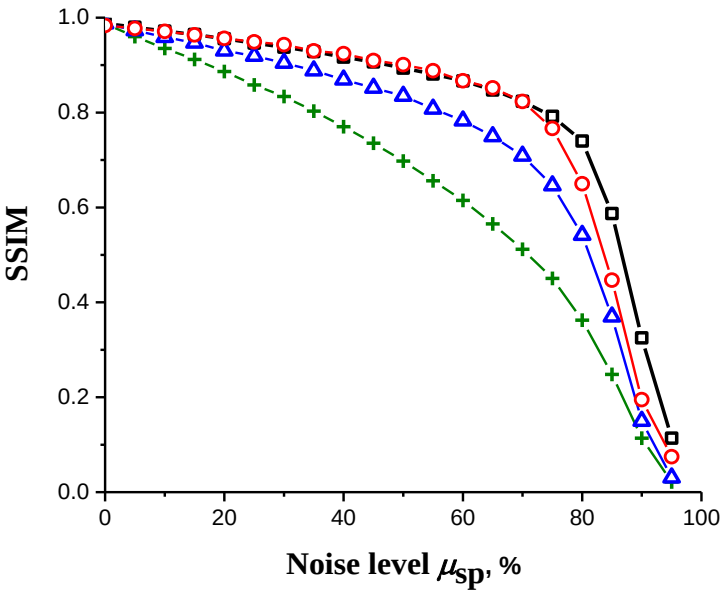

**Fig.8.** Dependence of SSIM metrics (SSIM-Img – black squares, SSIM-Map – circles, rectangles, and crosses) on the SP noise level $\mu_{sp}$ for recursive denoising by MF with different windows size of entropy maps: (7×7) px – red circles, (5×5) px – blue rectangles, (3×3) px – green crosses. MF parameters are in Table 1, the number of recursions is 10

The introduction of recursive processing (Fig. 5) has little effect on denoising efficiency at low noise levels. Indeed, within the initial noise $\mu_{sp}$ from zero to 10-15%, the SSIM values of Fig. 5a and 5b are almost indistinguishable from the case of single-pass filtration with thresholding (Fig. 4b). However, as $\mu_{sp}$ increases further, MF improves markedly. Up to very high noise ($\mu_{sp} \geq 70$ %) the median filter yields substantially higher SSIM-Img and SSIM-Map , especially with recursive thresholding after ten recursions, that is shown at Fig. 5b.

The functions SSIM-Img ($\mu_{sp}$) and SSIM-Map ($\mu_{sp}$) are well fitted by exponential sigmoids, which is shown by the nonlinear regression curves in Fig. 4 and 5. As the number of recursions increases (Fig. 5), the MF sigmoids become slightly broader and the noise level at which the curve crosses 0.5 (midpoint/inflection) shifts to the right relative to the curves in Fig. 4b. Thus, after ten threshold recursions (Fig. 5b), the MF is capable of restoring the image to a discernible quality even with a noise level above 80%. Note that SSIM-MAP curves are consistently below SSIM-Img at the same noise level.

Simulations show that further increasing the number of recursion steps does not improve denoising using MF. The metrics of restored image reach a steady-state level and do not change further. The dependencies of both SSIM metrics (SSIM-Img and SSIM-Map) on the threshold for MF (Fig. 6) exhibit a maximum, the threshold value of which is actually the same for different noise levels. Note that SSIM-Img maximum is slightly shifted to the right relative to the SSIM-Map maximum.

An important study is the analysis of the dependence of the SSIM metrics on the window size. Dependence of the SSIM metrics on the SP noise level for recursive denoising by MF with different filter windows size ((3×3), (5×5), (7×7), and (9×9) px) presented in Fig. 7. As can be seen, both metrics have almost identical dependence on noise at low and medium levels (with a slight advantage for window (3×3) px). However, for high noise levels, the filter's metrics with window (3×3) px drop sharply compared to noise reduction with larger windows.

Dependence of the SSIM-Map on the SP noise level with different windows size of entropy maps ((3×3), (5×5), and (7×7) px) in comparison of SSIM-Img for MF shown in Fig. 8. As can be seen, with the increase in the size of the entropy map windows, SSIM-Map curves close in the curve of the SSIM-Img metric. Note that the entropy map with window (3×3) px shows a curve that is very distant (towards lower values) from the SSIM-Img curve. This window will be used later to calculate image entropy maps.

At the end of this section, Figure 9 and 10 demonstrate some options for noise reduction in images with medium and high noise levels using recursive MF with thresholding. As can be seen, MF restores the image with medium noise level ($\mu_{sp}$ = 30 %) well, what is also shown by the SSIM metrics in Table 2. The MF with window (3×3) px restores the noisy image almost perfectly (Fig.9a). However, at the high noise level ($\mu_{sp}$ = 60%), noticeable differences are observed in the restored images also for MFs with different windows size. MF with window (3×3) px does not eliminate all the defects (Fig.9b). Only the MF with large window (5×5) px can effectively remove noise and restore the image (Fig. 10b) increasing the SSIM metrics (SSIM-Img and SSIM-Map) compared to window (3×3) px by 3 % (Table 2). It can also be noted that the point defects on the entropy maps (with entropy windows size (3×3) px) of Fig. 9 and 10 (bottom) are more noticeable than on the real images.

**Table 2.** SSIM metrics of restored images and their entropy maps using recursive MF-based schemes (Fig. 9, 10, 12)

| SSIM | MF, $W_1$=(3×3) px Fig.9 | | MF, $W_2$=(5×5) px Fig.10 | | 2-SRMAT Fig.12 | |
|---|---|---|---|---|---|---|
| | (a) | (b) | (a) | (b) | (a) | (b) |
| SSIM-Img | 0.955 | 0.834 | 0.942 | 0.863 | 0.945 | 0.868 |
| SSIM-Map | 0.844 | 0.522 | 0.831 | 0.539 | 0.837 | 0.565 |

**MF W1 = (3×3) px**

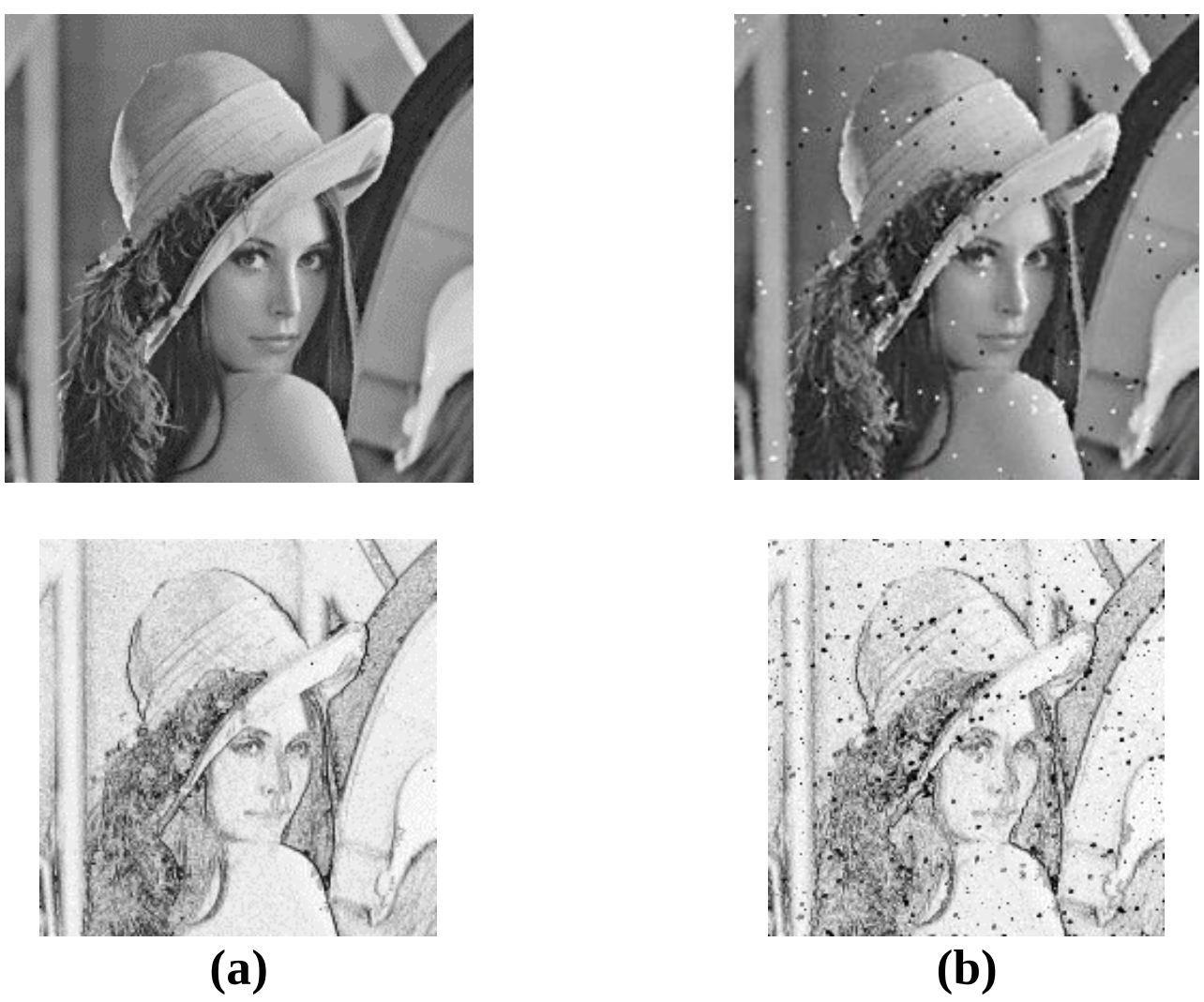

**Fig. 9.** Restored images (up) and their entropy maps (bottom) obtained using MF with windows size (3×3) px for the SP corrupted images: (a) $\mu_{sp}$ =30 %; (b) $\mu_{sp}$= 60 %. SSIM metrics are in Table 2. Entropy windows size is (3×3) px, MF calculation parameters (except for the window size) are in Table 1, the number of recursions is 20

**MF W1 = (5×5) px**

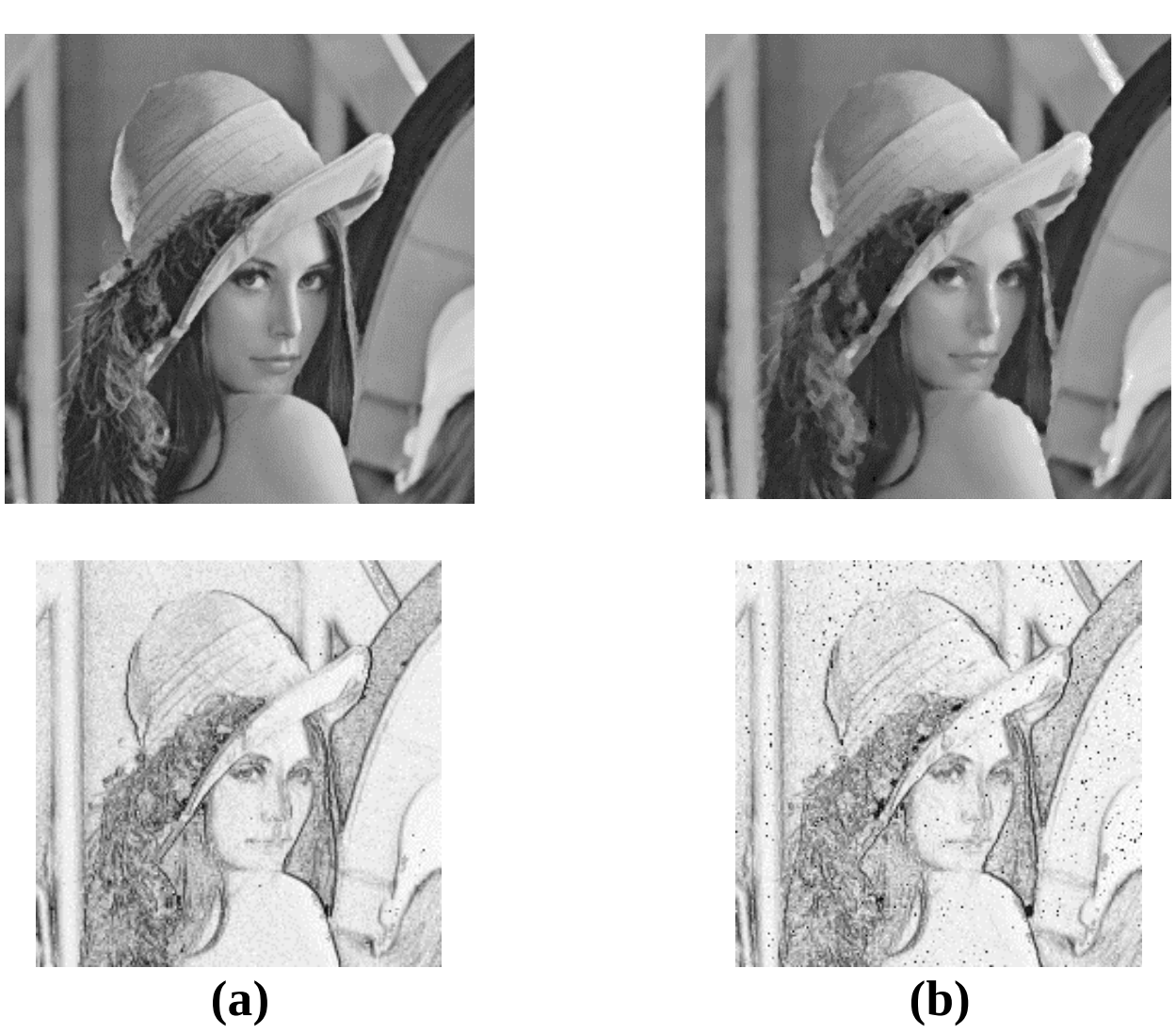

**Fig. 10.** Restored images (up) and their entropy maps (bottom) obtained using MF with windows size (5×5) px for the SP corrupted images: (a) $\mu_{sp}$ = 30 %; (b) $\mu_{sp}$ = 60 %. SSIM metrics are in Table 2. Entropy windows size is (3×3) px, MF calculation parameters are in Table 1, the number of recursions is 20

### 3.2 SP removal based on the 2-SRMAT scheme

As can be seen in Fig. 9b, the pixels in the restored image by the MF with windows (3×3) px, which were initially scattered uniformly (see, Fig. 2c), are combined into "islands". These clusters of defects cannot be removed merely by increasing the number of recursions or changing the threshold value while keeping the same window size. However, increasing the window size to (5×5) px eliminates the islands — the output image contains virtually no impulse corruptions (Fig. 10b,up). The rare small inclusions of damaged pixels can be seen only on the entropy map (Fig. 10b, bottom).

**Table 3.** Parameters of the 2-SRMAT scheme (Fig.11)

| **2-SRMAT** | Recursions | Step 1, threshold_1 | Step 2, threshold_2 |
|---|---|---|---|
| MF W1, (3×3) px | 20 | 0.15 | 0.15 |
| MF W2, (5×5) px | | | |

The effect of residual noise is well known, and various methods are used to eliminate it, primarily adaptive filtering [8]. We will examine this problem from the perspective of SSIM image analysis and entropy maps, using windows of varying sizes (without adaptation), i.e., MF scaling.

As noted in the introduction, with an increase in the kernel (window) size, noise reduction is accompanied by increased blurring, and this can be seen in the comparison of Fig. 9b and 10b. Indeed, the restored images by the MF with windows (3×3) px (Fig. 9b, left) does not provide complete noise reduction, but it appears slightly higher in contrast than that the image obtained with windows (5×5) px (Fig. 10b). This is especially noticeable at lower resolution images, increasing their size.Thus, increasing the window size provides more effective noise reduction, including the removal of island defects, but at the expense of additional blurring.

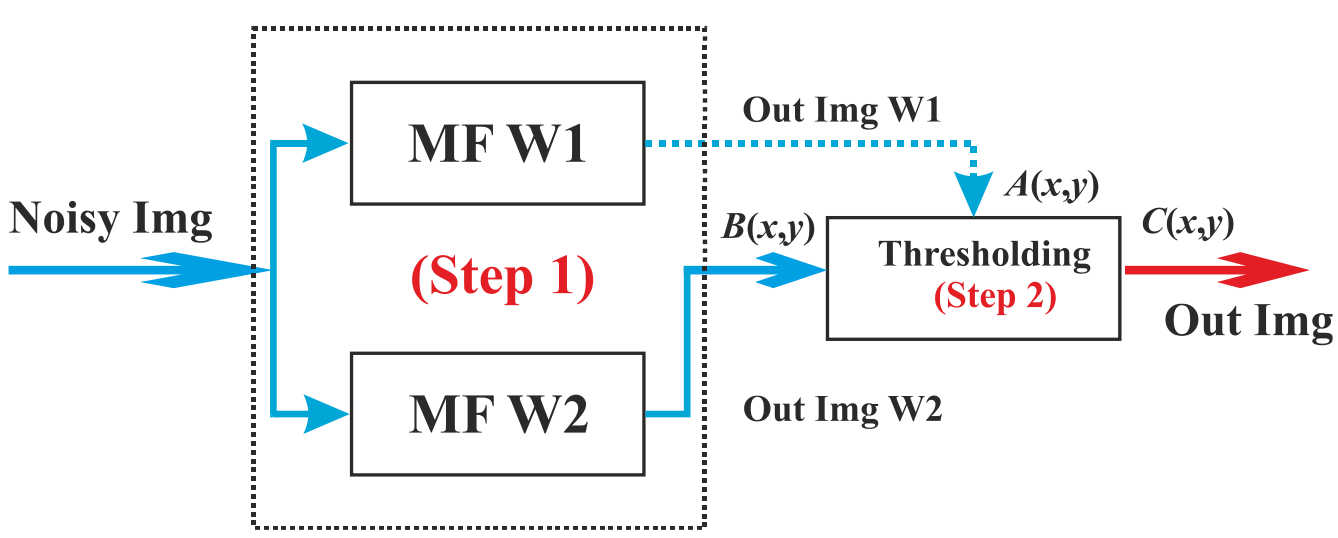

**Fig. 11.** 2MFs denoising scheme (2-SRMAT) for SP corrupted images

**2-SRMAT**

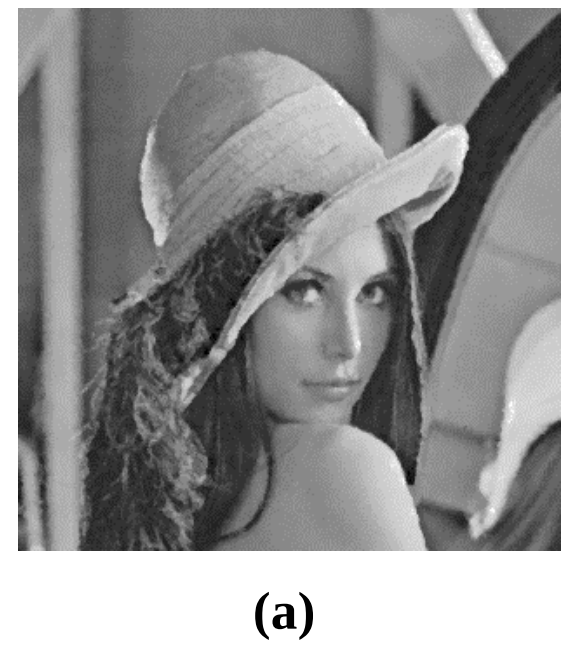

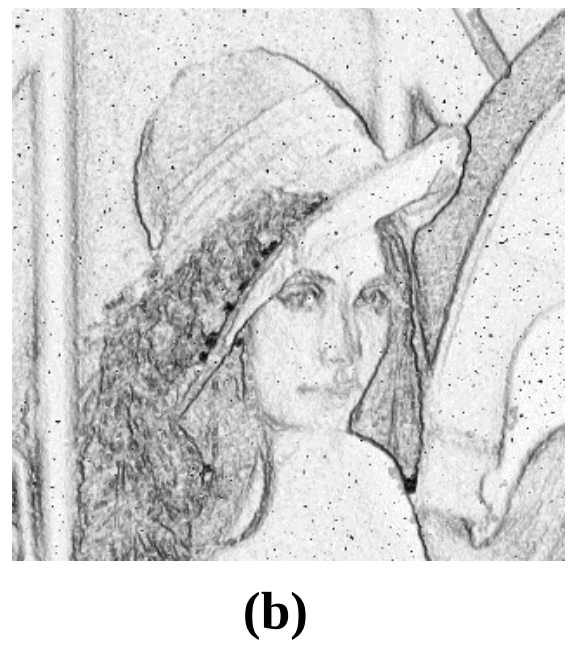

**(a)** **(b)**

**Fig. 12** Restored images (a) and their entropy maps (b) obtained using 2MFs scheme (Fig. 11) – 2-SRMAT for one SP noisy image with $\mu_{sp}$ = 60 %. SSIM metrics are in Table 2. Entropy windows size is (3×3) px, parameters of the 2-SRMAT scheme are in Table 3

We propose a promising two-scale scheme that combines two MFs (Fig. 11) with two different windows based on thresholding (2). This algorithm is in two steps:
***Step 1.*** Apply recursive threshold denoising (multiple passes) to the noisy image (Noisy Img) with two independent MFs using windows $W_1$ and $W_2$ ($W_1 < W_2$) as in Fig.1 and with the same threshold value (threshold_1), producing the output images Out Img W1 and Out Img W2, respectively.
***Step 2*** –Thresholding**.** Use rule (2) with threshold_2, where $A(x, y)$ is Out Img W1, $B(x, y)$ is Out Img W2, and $C(x, y)$ is the final output image (Out Img).

We refer to this 2MFs scheme Fig. 11 as a two-level scaled recursive median algorithm with thresholding (2-SRMAT), since two window scales are combined. Fig. 12 shows the restoration of moderately and heavily SP damaged images by the 2-SRMAT with $W_1 = (3\times3)$ px and $W_2 = (5\times5)$ px and two thresholds at Steps 1 (threshold1) and 2 (threshold2). Based on the visual assessment of Fig. 12a, as well as the results of calculating the SSIM metrics (Table 2), it can be concluded that for the medium noise level ($\mu_{sp}$= 30 %), the combined 2-SRMAT scheme has no advantages over individual filters (Figs. 9a and 10a).

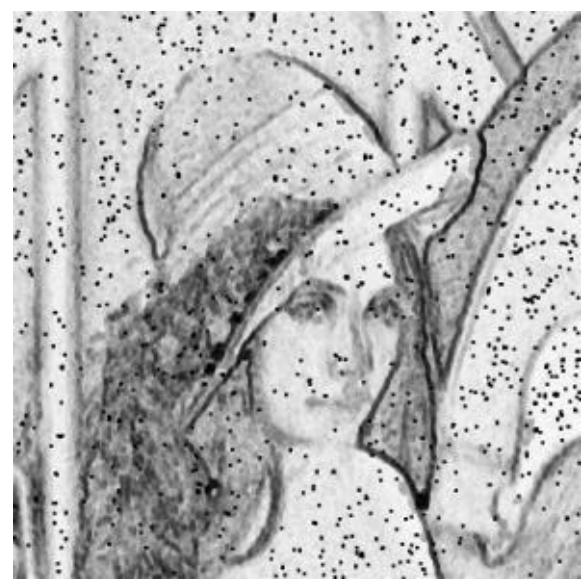

**Fig. 13.** Entropy maps of restored images using 2-SRMAT of Fig. 12b, transformed by the erosion method with kernel (3×3) px and one iteration (OpenCV, Python)

The reason is that the output images of MFs with windows W1 = (3 × 3) px and W2 = (5×5) px either contain very few island defects or do not contain them at all. Thus, the new image in the second step of 2-SRMAT cannot be improved by simultaneously preserving contrast and removing island defects from the image restored by the smaller-window MF. In contrast, denoising a highly noisy image using 2-SRMAT produces a visually improved result, suppressing most visible im pulse defects and improving the image obtained using individual filters. Table 2 shows that the SSIM-Img values of MF with W2 = (5×5) px and 2-SRMAT differ by 0.5%, while the difference between the SSIM-Map values of these filters is 2.6%.

Figure 13 shows the entropy map from Fig. 12b (2-SRMAT), morphologically eroded to enhance the contrast of island defects. The map contains numerous de fects, but they become clearly visible only after increasing the contrast. Thus, entropy maps are more sensitive to SP noise: SP defects are clearly visible in entropy maps, while they are virtually invisible in images (compare Figures 12a and 13).

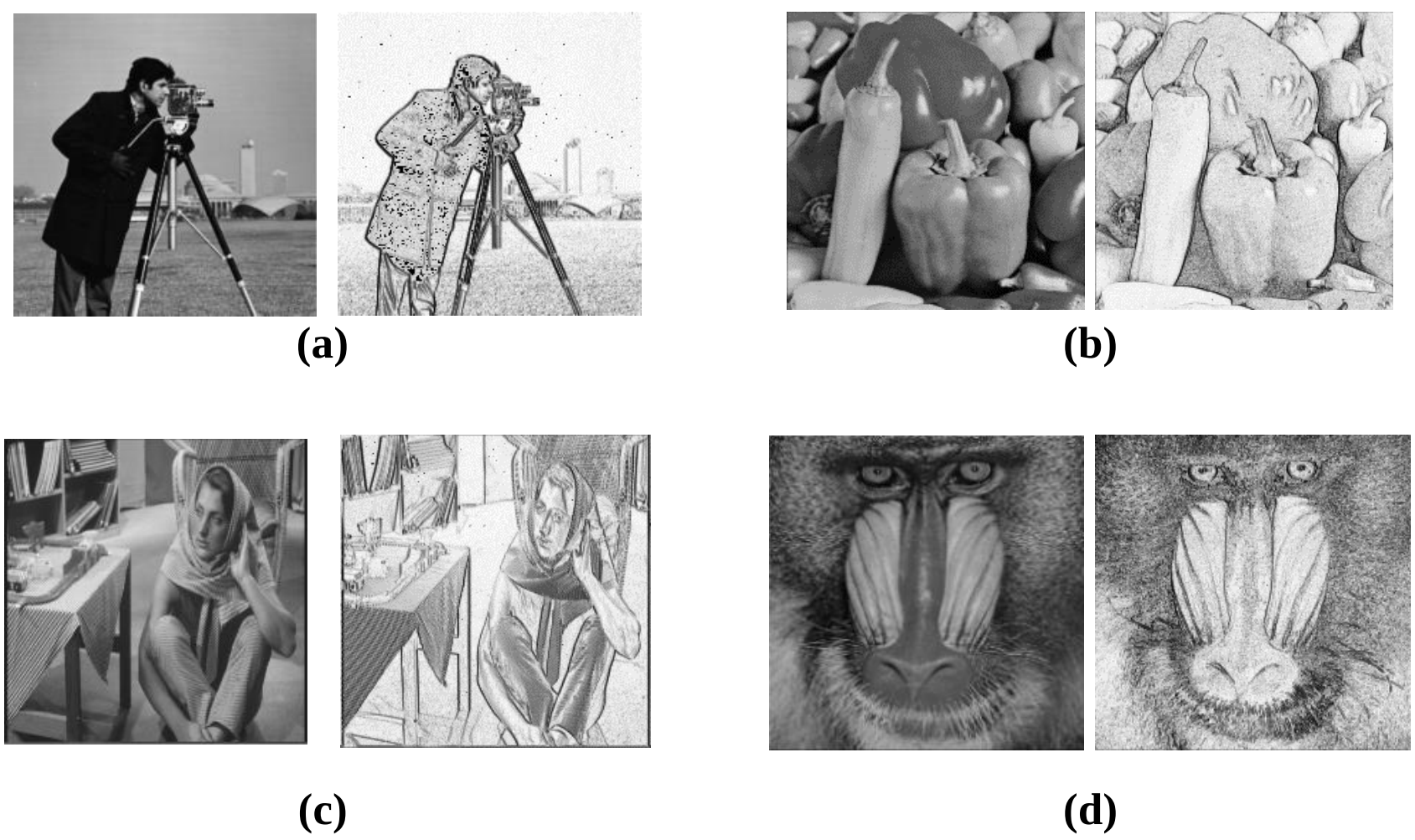

**Fig. 14.** Clean grayscale images of (512×512) px (left) and their 8-bit entropy maps (right) computed by singular value decomposition using sliding windows of (3×3) px: Cameraman (a), Peppers (b), Barbara (c), Monkey (d)

**Table 4.** Comparison of SSIM-Img, SSIM-Map and PSNR of restored grayscale images by the 2-SRMAT scheme for SP-corrupted images with $\mu_{sp} = 30 \div 70\%$. The parameters of the 2-SRMAT scheme are given in Table 3. Each denoising metric represents the average value of 10 tests

| Image/Metrics | Noise densities, % | | | | |
|---|---|---|---|---|---|
| | 30 | 40 | 50 | 60 | 70 |
| **Lena** | | | | | |
| SSIM_Img | 0.945 | 0.926 | 0.90 | 0.868 | 0.819 |
| SSIM_MAP | 0.837 | 0.771 | 0.68 | 0.565 | 0.40 |
| PSNR | 31.7 | 30.3 | 29.3 | 27.8 | 26.3 |
| **Cameraman** | | | | | |
| SSIM_Img | 0.924 | 0.901 | 0.875 | 0.843 | 0.795 |
| SSIM_MAP | 0.687 | 0.605 | 0.503 | 0.384 | 0.266 |
| PSNR | 29.9 | 28.6 | 27.3 | 26.2 | 24.3 |
| **Peppers** | | | | | |
| SSIM_Img | 0.921 | 0.897 | 0.869 | 0.832 | 0.782 |
| SSIM_MAP | 0.774 | 0.705 | 0.626 | 0.514 | 0.383 |
| PSNR | 28.4 | 27.8 | 27.1 | 26.0 | 24.7 |
| **Barbara** | | | | | |
| SSIM_Img | 0.910 | 0.883 | 0.851 | 0.812 | 0.744 |
| SSIM_MAP | 0.790 | 0.723 | 0.643 | 0.511 | 0.393 |
| PSNR | 28.9 | 27.5 | 26.7 | 25.4 | 23.4 |
| **Monkey** | | | | | |
| SSIM_Img | 0.819 | 0.779 | 0.732 | 0.666 | 0.578 |
| SSIM_MAP | 0.678 | 0.604 | 0.522 | 0.419 | 0.296 |
| PSNR | 24.5 | 23.8 | 23.2 | 22.2 | 21.2 |

Table 4 extends the evaluation from a single image and noise level to five standard grayscale images and five noise densities. Restoration quality depends strongly on image texture, contrast, and fine-detail content. Lena gives the highest values among the tested images, whereas Monkey gives the lowest values because fine hair-like structures are strongly affected by dense impulse corruption and median filtering. Cameraman is particularly informative: its SSIM-Img remains relatively high, while its SSIM-Map is much lower, consistent with the loss and blurring of local fine structure. Peppers and Barbara retain comparatively strong SSIM-Map values because their dominant contrast structures remain detectable after restoration.

**Table 5.** A comparison of SSIM-Img across different methods for the grayscale *Lena* image (512×512 px). Methods: detail-aware filter (DAF) [22], decision-based algorithm (DBA) [23], noise adaptive fuzzy switching median filter (NAFSM) [24], effective noise adaptive median filter (ENAMF) [25], different applied median filter (DAMF) [26], adaptive Riesz mean filter (ARmF) [27], deep convolutional neural network (DCNN) [28]. This work: 2-SRMAT – Fig.12a

| Method | DAF | DBA | NAFSM | ENAMF | DAMF | ARmF | **DCNN** | 2-SRMAT |
|---|---|---|---|---|---|---|---|---|
| SSIM-Img | 0.89 | 0.84 | 0.87 | 0.86 | 0.87 | 0.87 | **0.93** | 0.868 |

Finally, Table 5 reports denoising results for the grayscale *Lena* image (512×512 px) with fixed SP noise ($\mu_{sp}$ = 60 %) for methods [22–28], compared with the algrithms proposed in this study. The values for the compared methods are taken from the corresponding publications; therefore, minor differences in noise generation, pre processing, and SSIM implementation may affect direct comparability. This table should be understood as a contextual comparison rather than a complete state-of the-art benchmark. The results indicate that 2-SRMAT achieves SSIM-Img values comparable to several classical adaptive median-based methods, while remaining algorithmically simple and requiring no training stage. The DCNN method [28], highlighted in bold, uses a set of noisy Lena images for training and should therefore be considered separately from non-learning filtering schemes. Cascade decision-based schemes have also been proposed for high-density SP noise removal [29]. For this reason, 2-SRMAT should be evaluated not only by the final SSIM-Img value, but also by the absence of training, the number of tunable parameters, computational time, and the entropy-map response to residual artifacts.

## 4 Discussion

The simulations in Section 3.1 corroborate a known result [8,9]: as the number of recursions increases, the MF converges to a stationary root image that no longer changes and achieves maximal SSIM-Img and SSIM-Map.

A key concept of the proposed 2MF/2-SRMAT scheme is the combination of two median-filter outputs followed by a final thresholding step. The first output, obtained with a smaller window, better preserves local contrast but may still contain island like impulse artifacts. The second output, obtained with a larger window, is more blurred but is largely free of residual impulse corruption. In the final thresholding step, pixels that are likely to be corrupted in the small-window output are replaced by the corresponding pixels from the large-window output. As a result, the final image is less blurred than the large-window output while containing fewer residual SP artifacts than the small-window output.

The comparison in Table 4 shows that recursive median filtering with thresholding can achieve results comparable to several established SP-noise removal methods [22, 28]. The proposed 2-SRMAT scheme is not intended to outperform all state-of-the-art methods in terms of SSIM-Img alone. Rather, its advantage lies in its simplicity: it requires no training stage, no explicit damaged-pixel detection module, and no complex adaptive parameter selection. Although more complex approaches such as DAF [22] and DCNN [28] achieve higher SSIM-Img values, 2-SRMAT provides a favorable simplicity–quality trade-off and remains attractive for resource-constrained implementations. Accordingly, the method is presented as a transparent baseline for high-density impulse-noise removal combined with an entropy-map-based diagnostic metric, without overstating universal algorithmic superiority.

Table 6 clarifies the methodological scope of the paper. Previous studies support entropy as an informative image-quality descriptor and show that SSIM can be applied to transformed representations, such as wavelet or edge maps. However, the present work differs by explicitly constructing pixel-aligned entropy maps for the clean and restored images and then applying standard SSIM to these maps

**Table 6.** Positioning of the proposed SSIM-Map metric with respect to related entropy-based and SSIM approaches. The comparison is based on external literature; the last row describes the present work rather than a separate prior publication

| Approach | Entropy representation | SSIM / structural comparison | Objects compared | Relation to the present SSIM Map |
|---|---|---|---|---|
| Regional or scalar entropy IQA [30,31] | Regional or global entropy features | Entropy is used as a quality indicator or compared with conventional metrics | Scalar or regional quality descriptors | Useful background, but not SSIM between entropy maps |
| Blind entropy degradation IQA [32] | Local intensity, gradient, and orientation distributions | No full-reference entropy-map SSIM | Feature distributions for blind quality assessment | Entropy supports quality assessment, but not as a pixel aligned map compared by SSIM |
| Wavelet / edge/ high frequency SSIM [33,34] | No entropy map | SSIM is applied to transformed image content | Wavelet subbands, edges, or high-frequency regions | Conceptually close as SSIM on-representation, but uses different representations |
| Structure-texture IQA [35,36] | Texture / feature statistics; entropy may guide feature weighting | Feature-space structure and textures imilarity | CNN/statistical feature maps | No direct SSIM between entropy images |
| 2D entropy texture analysis [37-39] | Local 2D entropy/ irregularity measures | Usually not used as full-reference SSIM target | Texture or irregularity maps/features | Provides foundations for entropy maps, but not the final SSIM-Map step |
| This work | SVD-entropy map in sliding windows | SSIM is computed directly between entropy maps | Clean-reference entropy map vs. restored image entropy map | Proposed contribution: a full reference structural metricf or local irregularity preservation |

Entropy maps more strongly reveal subtle distortions that are virtually invisible in the images themselves, which is reflected in the sensitivity of SSIM-Map compared to SSIM-Img. In our opinion, the SSIM-Map metric as IQA is most useful for fine-tuning the evaluation of very similar images, when the SSIM-Img is close to one, and the SSIM-Map is even lower 0.85–0.9 (see, for example, Table 2, $\mu_{sp}$ = 30%). It is also important to calculate entropy maps with a small window, since in this case SSIM-Img and SSIM-Map differ significantly over a wide range of noise levels (see the green and black curves in Fig. 8).

There are many entropy measures, including two-dimensional variants, to construct entropy maps, and they vary in segmentation and texture detail. Depending on the entropy type and parameters, "entropy images" can be highly or poorly detailed, reflecting the metric's sensitivity to local intensity variations. We focused on SVDEn, which, on the one hand, produces a map with the clearest definition of key structures, for example, Lena's hat, hair, face, and eyes. On the other hand, SVDEn calculation is the fastest, as it relies not on statistics but on a linear decomposition of matrices.

## 5 Conclusion

This work analyzed recursive thresholded median filtering for the removal of SP noise in grayscale images using two complementary structural similarity measures: conventional image-level SSIM and entropy-map-based SSIM. The results show that recursive thresholding improves median-filter restoration at high impulse-noise densities, while SSIM-Map provides a more conservative assessment of residual local artifacts and blur. The proposed two-scale 2-SRMAT scheme combines the advantages of small and large median-filter windows: the smaller window helps preserve local contrast, whereas the larger window suppresses residual island-like impulse defects. The main practical implication is that entropy-map SSIM can be used as an additional diagnostic layer when conventional image-domain metrics give similar values for restorations with different residual artifacts or smoothing behavior. The resulting approach is simple, does not require training, and is potentially suitable for resource-constrained implementations. Future research should address adaptive threshold selection, broader multi-image benchmarking, mixed-noise and color-image scenarios, the use of alternative entropy measures for constructing quality-sensitive image maps, as well as the expansion of the SSIM-Map application scope (recognition, classification, compression, coding, etc.).

**Authors Contributions**

Petr Boriskov, conceptualization, methodology and software; Petr Boriskov and Andrei Velichko, investigation, revised and finalized the paper; Andrei Velichko, drafted the manuscript and project administration

**ORCID**

Petr Boriskov - http://orcid.org/0000-0002-2904-9612

Andrei Velichko - http://orcid.org/0000-0002-9341-1831